\documentclass{osa-article}

\journal{oe}

\usepackage{amsmath}  
\usepackage{amssymb}
\usepackage{gensymb}
\usepackage{braket}

\usepackage{amsfonts} 
\usepackage{graphicx} 

\usepackage{color}

\usepackage{comment}
\usepackage{dcolumn}
\usepackage{bm}
\newcommand{\Note}[1]{\textcolor{black}{ #1}}

\articletype{Research Article}

\begin{document}

\title{A high-speed, wavelength invariant, single-pixel wavefront sensor with a digital micromirror device}

\author{Mitchell~A.~Cox,\authormark{1,*} Ermes~Toninelli,\authormark{2,*} Ling~Cheng,\authormark{1} Miles~Padgett,\authormark{2} and Andrew~Forbes\authormark{3,$\dagger$}}

\address{\authormark{1}School of Electrical and Information Engineering, University of the Witwatersrand, Johannesburg, ZA\\
\authormark{2}SUPA, School of Physics and Astronomy, University of Glasgow, Glasgow, G12 8QQ, UK\\
\authormark{3}School of Physics, University of the Witwatersrand, Johannesburg, ZA}

\email{\authormark{*}these authors contributed equally to the paper, \authormark{$\dagger$}andrew.forbes@wits.ac.za} 



\begin{abstract}
The wavefront measurement of a light beam is a complex task, which often requires a series of spatially resolved intensity measurements. For instance, a detector array may be used to measure the local phase gradient in the transverse plane of the unknown laser beam.  In most cases the resolution of the reconstructed wavefront is determined by the resolution of the detector, which in the infrared case is severely limited.  Here we employ a Digital Micro-mirror Device (DMD) and a single-pixel detector (i.e. with no spatial resolution) to demonstrate the reconstruction of unknown wavefronts with excellent resolution.  Our approach exploits modal decomposition of the incoming field by the DMD, enabling wavefront measurements at 4~kHz of both visible and infrared laser beams.
\end{abstract}

\section{Introduction}


Wavefront sensing, preservation and/or correction \Note{is essential in many optical systems, including in astronomy with low intensity point-like sources of rays, tightly focussed medium-intensity laser beams in microscopy and imaging, and for the delivery without aberrations of high-power laser beams for materials processing \cite{marois_DirectImagingMultiple_2008,rueckel2006adaptive,cizmar_SituWavefrontCorrection_2010,mauch2012adaptive}. Implicit in this is the understanding that most optical processes are phase rather than intensity dominant, thus phase and wavefront knowledge is paramount \cite{soldevila2018phase}.} \Note{It may be useful to point out that unlike object reconstruction by digital holography \cite{park2018quantitative} or computational imaging \cite{edgar2018principles}, here there is no object, no structured illumination, and no reference beam - it is the primary beam itself that must be probed and analysed by some in-line and preferably real-time device.  Often the outcome of such a wavefront measurement is a means to correct it, perhaps by adaptive optics.} Such wavefront sensing techniques rely on the ability to measure the phase of light which can only be indirectly inferred from intensity measurements. Methods to do so include ray tracing schemes, intensity measurements at several positions along the beam path, pyramid sensors, interferometric approaches, computational approaches, the use of non-linear optics, computer generated holograms (CGHs), meta-materials and polarimetry \cite{navarro1999laser,almoro2006complete,chamot2006adaptive,velghe2005wave,yang2018generalized,bruning2013comparative,huang2015real,borrego2011wavefront,kamali2018review,dudley2014all,Ruelas2018,changhai2011performance,shin2018reference,baek2018high}.  Perhaps the most well-known is the Shack-Hartmann wavefront sensor \cite{lane1992wave,vohnsen2018hartmann}.  Its popularity stems from the simplicity of the configuration as well as the fact that the output can easily be used to drive an adaptive optical loop for wavefront correction.  More recently a modal approach to beam analysis has been demonstrated \cite{flamm2013all,litvin2012azimuthal,schulze2012wavefront,schulze2013reconstruction,litvin2011poynting,schulze2013measurement,liu2013free,godin2014reconstruction}. Using both hard-coded CGHs and digital holograms on spatial light modulators (SLMs) (see \cite{forbes2016creation} for a review), the technique was shown to be highly versatile and accurate.  These approaches to wavefront-sensing and corrections still suffer from slow refresh rates, often limited to 100s of Hz, are usually expensive (especially for non-visible applications), and are limited both in terms of spatial resolution and operational wavelength-range.

In this work we demonstrate a wavefront-sensor that is broadband (spanning over 1000 nm, from the visible to the mid-IR), fast (with a refresh rate in the kHz range), and inexpensive (100s of US dollars). We achieve this by building our wavefront-sensor around a digital micro-mirror device (DMD) and leveraging the advantages of the modal decomposition technique. This enables the rapid production of reconstructed intensity and phase-maps with an ``unlimited'' resolution, even though the employed detector is a single-pixel ``bucket-detector''. \Note{We demonstrate the technique using both a visible and NIR laser programmatically deteriorated with aberrations typical of moderately distorted beams, e.g., as would be experienced with thermally distorted high-power laser beams, propagation through a moderately turbulent atmosphere, and optically distorted beams due to tight focusing or large apertures. We demonstrated excellent wavefront reconstruction with measurement rates of 4000 Hz, fast enough to be considered real-time for most practical applications.}

\section{Background theory}
\label{sec:theory}
\noindent For the aid of the reader we briefly introduce the notion of wavefront and phase, outlining how it may be extracted by a modal decomposition approach. 
\subsection{Wavefront and phase}
\noindent The wavefront of an optical field is defined as the continuous surface that is normal to the time average direction of energy propagation, i.e., normal to the time average Poynting vector $\mathbf{P}$ 
\begin{equation}
w(\mathbf{r},z)\perp\mathbf{P}(\mathbf{s},z),
\label{eq:wf1}
\end{equation}
where $z$ denotes the position of the measurement plane. The ISO standards define the wavefront more generally as the continuous surface that minimizes the power density weighted deviations of the direction of its normal vectors to the direction of energy flow in the measurement plane 
\begin{equation}
\int\int|\mathbf{P}|\left|\frac{\mathbf{P}_t}{|\mathbf{P}|}-\nabla_t w\right|^2dA\rightarrow\mathrm{min},
\label{eq:wf2}
\end{equation}
where $\mathbf{P}_t=[P_x,\,P_y,\,0]'$. What remains then is to find the Poynting vector $\mathbf{P}$; this is computable from the knowledge of the optical field by
\begin{equation}
\mathbf{P}(\mathbf{s})=\frac{1}{2}\Re{\left[\frac{i}{\omega\epsilon_0}\epsilon^{-1}(\mathbf{s})[\nabla\times\mathbf{U}(\mathbf{s})]\times\mathbf{U}^\ast(\mathbf{s})\right]},
\label{eq:poynting1wf}
\end{equation}
where $\Re$ denotes the real component, for vector fields $\mathbf{U}$, and by
\begin{equation}
\mathbf{P}(\mathbf{s})=\frac{\epsilon_0\omega}{4}\left[i(U\nabla U^\ast-U^\ast\nabla U)+2k|U|^2\mathbf{e}_z\right]
\label{eq:poynting2wf}
\end{equation}
for scalar fields $U$, where $\omega$ is the angular frequency, $\epsilon_0$ the vacuum permittivity, $\epsilon$ the permittivity distribution.  In the simple case of scalar, i.e. linearly polarized beams, the wavefront is equal to the phase distribution $\Phi(\mathbf{s})$ of the beam except for a proportionality factor
\begin{equation}
w(\mathbf{s})=\frac{\lambda}{2\pi}\Phi(\mathbf{s}) = \frac{\lambda}{2\pi}\text{arg}\{U(\mathbf{s})\},
\label{eq:wf3}
\end{equation}
where $\lambda$ is the wavelength. It is important to note that this expression is only valid so long as there are no phase jumps or phase singularities, because the wavefront is always considered to be a continuous surface.  Nevertheless, this facilitates easy extract of the wavefront by a phase measurement.  

From these expressions it is clear that if the optical field is completely known then the wavefront may readily be inferred.  Here we outline how to do this by a modal expansion into a known basis, commonly referred to as modal decomposition.

\subsection{Modal decomposition}
\label{sec:modalDecomp}

\noindent Any unknown field, $U(\mathbf{s})$, can be written in terms of an orthonormal basis set, $\Psi_n(\mathbf{s})$, 
\begin{equation}
\label{eq:U}
U(\mathbf{s}) = \sum_{n=1}^{\infty} c_n \Psi_n(\mathbf{s}) = \sum_{n=1}^{\infty} |c_n| e^{i\phi_n} \Psi_n(\mathbf{s}),
\end{equation}
with complex weights $c_n = |c_n| e^{i\phi_n}$ where $|c_n|^2$ is the power in mode $\Psi_n(\mathbf{s})$ and $\phi_n$ is the inter-modal phase, satisfying $ \sum_{n=1}^{\infty} |c_n|^2 = 1.$  Thus, if the complex coefficients can be found then the optical field and its wavefront can be reconstructed, usually requiring only a small number of measurements, especially in the case of common aberrations. Note that the resolution at which the wavefront may be inferred is not determined by the resolution of the detector. In other words, whereas only a few complex numbers are measured, the reconstructed resolution is determined by the resolution of the basis functions, which are purely computational. 

The unknown modal coefficients, $c_n$, can be found by the inner product 
\begin{equation}
\label{eq:mdoverlap}
c_n = \braket{\Psi_n|U} = \int \Psi_n^*(\mathbf{s}) U(\mathbf{s}) d\mathbf{s},
\end{equation}
\noindent where we have exploited the ortho-normality of the basis, namely
\begin{equation}
\braket{\Psi_n|\Psi_m} = \int \Psi_n^*(\mathbf{s}) \Psi_m(\mathbf{s}) d\mathbf{s} = \delta_{nm}.
\end{equation}

This may be achieved experimentally using a lens to execute an optical Fourier transform, $\mathfrak{F}$.  Accordingly we apply the convolution theorem
\begin{equation}
\label{eq:convtheorem}
\mathfrak{F}\{f(\mathbf{s})g(\mathbf{s})\} = F(\mathbf{k}) * G(\mathbf{k}) = \int F(\mathbf{k})G(\mathbf{s}-\mathbf{k}) d\mathbf{k}
\end{equation}

\noindent to the product of the incoming field modulated with a transmission function, $T_n(\mathbf{s})$, that is the conjugate of the basis function, namely,
\begin{equation}
W_0(\mathbf{s}) = T_n(\mathbf{s}) U(\mathbf{s}) = \Psi_n^*(\mathbf{s}) U(\mathbf{s}),
\end{equation}

\noindent to find the new field at the focal plane of the lens as 
\begin{equation}
\label{eq:mdlensft}
W_f(\mathbf{s}) = A_0~\mathfrak{F} \{W_0(\mathbf{s}) \} = A_0 \int \Psi_n^*(\mathbf{k}) U(\mathbf{s} - \mathbf{k}) d\mathbf{k}
\end{equation}
Here $A_0 = \exp(i4\pi f/ \lambda)/(i\lambda f)$ where $f$ is the focal length of the lens and $\lambda$ the wavelength of the light. If we set $\mathbf{s} = \mathbf{0}$, which experimentally is the on-axis (origin) intensity in the Fourier plane, then Eq.~(\ref{eq:mdlensft}) becomes
\begin{equation}
W_f(\mathbf{0}) = A_0 \int \Psi_n^*(\mathbf{k}) U(\mathbf{k}) d\mathbf{k}
\end{equation}
\noindent which is the desired inner product of Eq.~(\ref{eq:mdoverlap}).  Therefore we can find our modal weightings from an intensity measurement of 
\begin{equation}
I_n = |W_f(\mathbf{0})|^2 = |A_0|^2 |\braket{\Psi_n | U}|^2 = |c_n|^2.
\end{equation}

This is not yet sufficient to reconstruct the wavefront of the field as the inter-modal phases are also needed. The inter-modal phases $\Delta \phi_n$ for the modes $\Psi_n$ cannot be measured directly, however, it is possible to calculate them in relation to an arbitrary reference mode $\Psi_{\mathrm{ref}}$. This is achieved with two additional measurements, in which the unknown field is overlapped with the superposition of the basis functions \cite{flamm2009,schulze2013reconstruction}, effectively extracting the relative phases from the interference of the modes. Thus, in addition to performing a modal decomposition with a set of pure basis functions, $\Psi_n$, we perform an additional modal decomposition with each mode and a reference, described by the transmission functions
\begin{equation}
\label{eq:Tcos}
T^{\mathrm{cos}}_n (\mathbf{s}) = \frac{\left[ \Psi^*_{\mathrm{ref}}(\mathbf{s}) + \Psi^*_{n}(\mathbf{s}) \right]}{\sqrt{2}}
\end{equation}
\noindent and
\begin{equation}
\label{eq:Tsin}
T^{\mathrm{sin}}_n (\mathbf{s}) = \frac{\left[ \Psi^*_{\mathrm{ref}}(\mathbf{s}) + i\Psi^*_{n}(\mathbf{s}) \right]}{\sqrt{2}}.
\end{equation}
It is worth noting that, while in principle one measurement is sufficient for an inter-modal phase, two ensure that the phase value is not ambiguous. If the resulting intensity measurements are $I^{\mathrm{cos}}_n$ and $I^{\mathrm{sin}}_n$, then the inter-modal phase can be found from 
\begin{equation}
\label{eq:intermodalPhase}
\Delta \phi_n = - \arctan \left[ \frac{2I^{\mathrm{sin}}_n - I_n - I_{\mathrm{ref}}}{2I^{\mathrm{cos}}_n - I_n - I_{\mathrm{ref}}} \right] \in [-\pi, \pi].
\end{equation}
Importantly, in order to reduce the error in the estimation of the inter-modal phase, the reference mode should return an intensity comparatively high to the average intensity of the other modes in the basis.

In the present context, the transmission functions are implemented as computer generated holograms (CGHs), and displayed on a DMD spatial light modulator.  As a note, the amplitudes of the respective transmission functions are normalized to satisfy the condition that the encoded transmission function, $\widetilde{T_n}$, is $|\widetilde{T}_n| \in [0,1]$. As a result, generated or detected modes are still orthogonal but are no longer orthonormal, with deleterious effects for modal decomposition \cite{flamm2013all}.  It has been shown that it is paramount to re-scale the measured intensities before normalising the measurements for $\sum_n I_n = 1$ \cite{flamm2013all}. This correction must be done for each CGH in the system by simply multiplying in the additional factors, with the equations below for a single CGH:
\begin{equation}
\label{eq:orthonormCA}
I_n = I_{\mathrm{meas.}} \braket{\widetilde{T}_n^{\mathrm{CA}}|\widetilde{T}_n^{\mathrm{CA}}}^{-1} 
\end{equation}
\begin{equation}
\label{eq:orthonormPO}
I_n = I_{\mathrm{meas.}} |\widetilde{T}_n^{\mathrm{PO}}|^{-1}
\end{equation}
\noindent where $I_{\mathrm{meas.}}$ is the measured intensity which is re-scaled to result in $I_n$, depending on whether a Complex-Amplitude (CA) or a Phase Only (PO) CGH is used.\\

In order to encode the phase and amplitude of the desired transmission functions for implementation with a binary amplitude DMD, the following conditioning of the hologram is required~\cite{brown1966, lee1979} 
%
\begin{equation}
\label{eq:Tdmd}
\widetilde{T}_n(\mathbf{s}) = \frac{1}{2} + \frac{1}{2} \mathrm{sign} \left[ \cos{(p(\mathbf{s}))} + \cos(q(\mathbf{s})) \right],
\end{equation}
where
\begin{equation}
p(\mathbf{s}) = \arg (T_n (\mathbf{s})) + \phi_g (\mathbf{s})
\end{equation}
\begin{equation}
q(\mathbf{s}) = \arcsin\!\!\left(\frac{|T_n(\mathbf{s})|}{|T_n(\mathbf{s})|_{max}}\right)
\end{equation}
and $T_n$ is the desired function to be encoded (for example Eqs.~(\ref{eq:Tcos}), (\ref{eq:Tsin}) and (\ref{eq:LG})) and $\phi_g$ is a linear phase ramp which defines the period and angle of the resulting grating. The target field will occur in the first order diffraction spot. \Note{Due to the nature of a binary amplitude-only hologram, the efficiency is low in comparison to a phase-only hologram on a SLM. Efficiencies on the order of 1.5\% are expected, but this issue can be mitigated by using a sensitive detector, or seen as a benefit if higher incoming laser powers are expected \cite{Mirhosseini2013}.}\\

In this work we use the Laguerre-Gaussian (LG) basis as our expansion with basis functions in two indices given as \cite{Kogelnik1966}
\begin{equation}
\label{eq:LG}
\begin{aligned}
\Psi^{\mathrm{LG}}_{p,\ell}(r,\theta) = \sqrt{\frac{2p!}{\pi(p + |\ell|)!}} 
\left(\frac{r\sqrt{2}}{w_0} \right)^{|\ell|} 
\!\!L_p^{|\ell|} \!\!\left(\frac{2r^2}{w^2_0}\right)
\exp \!\!\left(-\frac{r^2}{w^2_0} \right)
\exp (-i\ell\theta)
\end{aligned}
\end{equation}
\noindent where $w_0$ is the Gaussian beam waist and $L_p^{|\ell|} (\cdot)$ is the generalised Laguerre polynomial with azimuthal index $\ell$ and radial index $p$. While the choice of basis is arbitrary there is always an optimal basis to minimise the number of modes in the expansion.  For example, if the measured mode has a rectangular shape then it is likely that the Hermite-Gaussian basis will be more suitable as it will require fewer terms in Eq.~(\ref{eq:U}) for an accurate reconstruction. 

\section{Experimental setup and methodology}
\label{sec:expsetup}

\begin{figure}[t]
\centering
\includegraphics[width=0.88\linewidth]{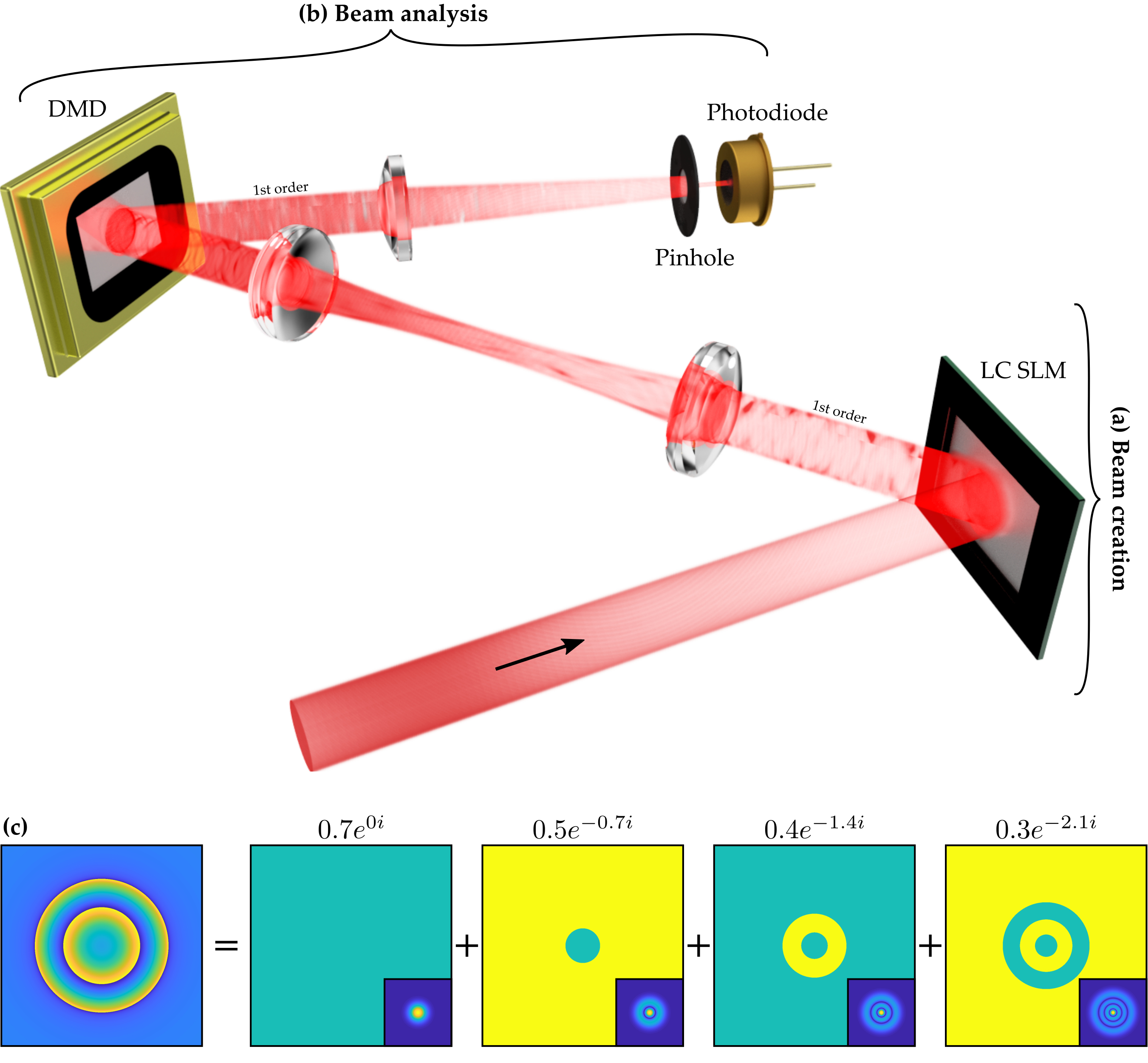}
\caption{Schematic representation of the experimental setup showing (a) mode (aberration) creation using a SLM and (b) modal decomposition using a DMD. When used as a wavefront measurement tool, part (a) would not be present and the incoming beam would shine directly onto the DMD. As an illustrative example, a modal decomposition of a defocus aberration to $\mathrm{LG}_{\ell=0}^{p\in[0,3]}$ is shown in (c), with modal weightings above each mode.}
\label{fig:expSetup}
\end{figure}

A schematic of the experimental setup is shown in Fig.~\ref{fig:expSetup}. We show a modal decomposition set-up which includes a DMD to display the CGH (the transmission function in Sec.~\ref{sec:theory}), a Fourier lens and a pinhole with a photo-detector to measure the on-axis intensity for the inner product outcome.  In order to select the on-axis mode for the modal decomposition, the photodiode can be either fibre-coupled (using a single-mode fibre) or paired with a precision pin-hole ($5~\mu$m).


In this work we tested two DMD devices.  The first, a DLP6500FYE device ($1920\times1080$ mirrors, $6.5~\mu$m pitch, and a refresh rate of 9.5~kHz), whose larger chip is on the one hand useful in displaying high order modes, but on the other hand is more affected by strain-induced curvature of the micromirror chip. Consequently, the results in this paper were primarily produced using the second device, a DLP3000, due to its smaller and thus optically flatter chip. This model has $608\times684$ mirrors (7.6~$\mu$m pitch, arranged in a diamond pattern) and a refresh rate of 4~kHz when switching through on-board memory patterns.  


We imposed a known primary aberration onto an incoming Gaussian beam and directed it towards the DMD wavefront sensor. For tests in the visible ($\lambda = 635$ nm) a camera was used as the detector and the intensity at origin (``single pixel'') used, while for the NIR ($\lambda = 1550$ nm) a single mode fibre coupled InGaS photodiode was used. A custom trans-impedance amplifier converted the photodiode current into a voltage that was then measured  by the 12~bit Analogue-to-Digital Converter (ADC) of an Arduino Due microcontroller, and sent to a computer. In order to operate the DMD at its fastest rate, the holograms were loaded onto its on-board flash memory.

\section{Reconstruction results}
\label{sec:results}

\subsection{Modal decomposition verification}
\label{subsec:verify}

\begin{figure}[t]
\centering
\includegraphics[width=1\linewidth]{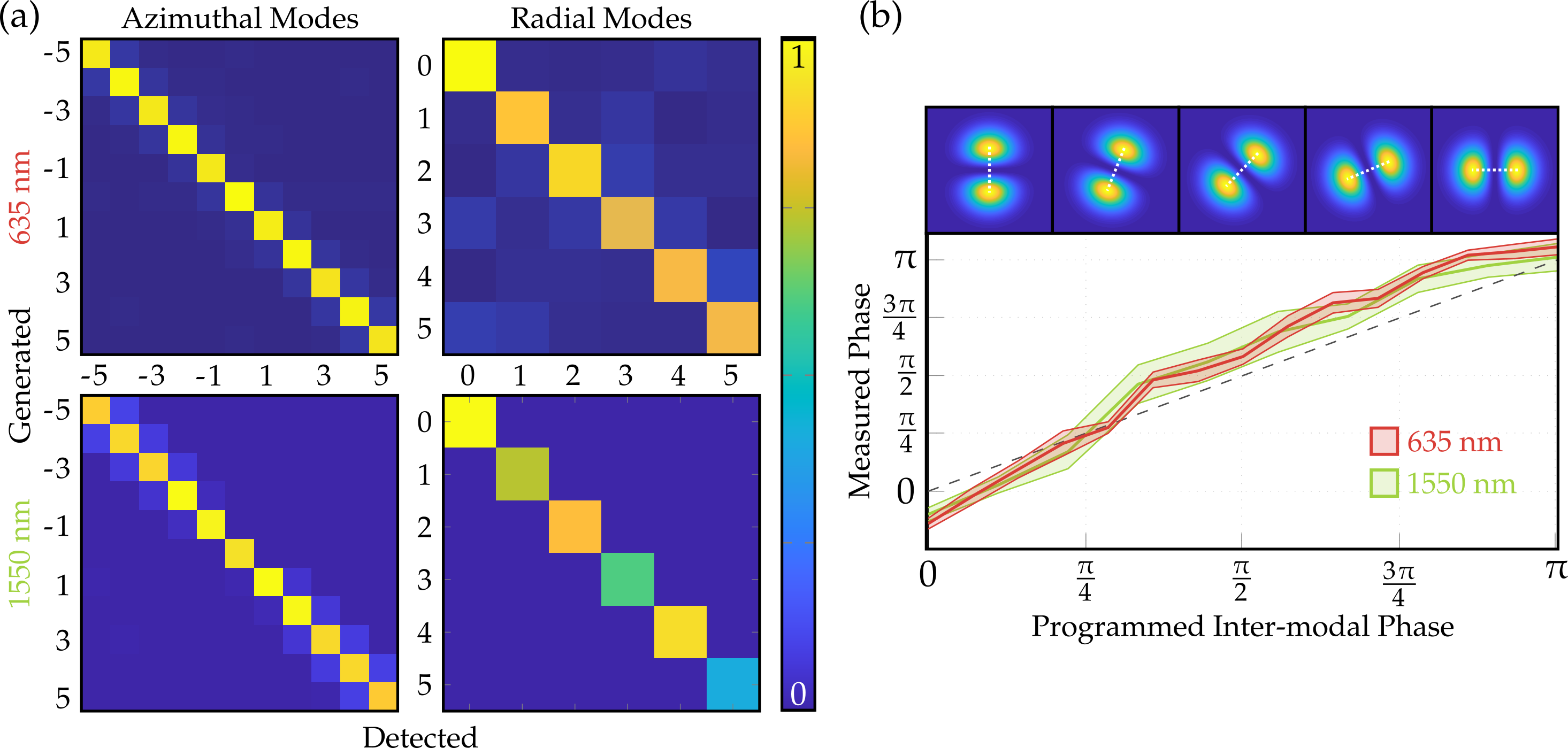}
\caption{(a) Experimental modal decomposition verification of modal amplitudes, $|c_n|$, of the experimental setup where each mode is generated and subsequently detected for both azimuthal ($\ell$) and radial ($p$) modes. (b) Verification of the inter-modal phase measurement, $\phi_n$, where a superposition of $\mathrm{LG}_{\ell=\pm 1}^{p=0}$ with a specific inter-modal phase was programmed and measured for both wavelengths. The slight crosstalk and phase error is caused by deformations of the DMD surface. }
\label{fig:sanityTests}
\end{figure}

In order to verify our wavefront sensor, a modal decomposition was performed on prepared Laguerre-Gaussian modes with $\ell\in[-3,3]$ and $p\in[0,3]$. Each mode was generated and a modal decomposition was performed for modal weights and inter-modal phases, with the results shown in Figs.~\ref{fig:sanityTests}. As expected, the azimuthal modal decomposition ($p=0$, $\ell\in[-5,5]$) at both wavelengths shows limited crosstalk, and thus a relatively linear measured intensity. 

The inter-modal phase measurement was verified by generating beam made from a superposition of two LG$_{\ell=\pm1}^{p=0}$ modes with a known phase shift between them, as in Eq.~(\ref{eq:lgpm1super}). The reference mode was chosen as the $\ell=-1$ mode 
\begin{equation}
\label{eq:lgpm1super}
T_n(\mathbf{s}) = \Psi^{\mathrm{LG}}_{\ell=-1}(\mathbf{s}) + e^{i\phi} \Psi^{\mathrm{LG}}_{\ell=1}(\mathbf{s}),
\end{equation}
where $T_n$ is the encoded transmission function and $\phi$ is the programmed inter-modal phase between the two modes. 

As shown in Fig.~\ref{fig:sanityTests}, both the visible and NIR inter-modal phase tests are largely correct within experimental error. Here, the measurements were repeated ten times as the phase reconstruction was found to be sensitive to noise, as indicated by the shaded error regions in the figure. The error for the NIR measurements was found to be negatively affected by the performance of our custom transimpedance amplifier used to sample the intensities from the photodiode.

\subsection{Wavefront measurements}
\label{subsec:results}
\begin{figure}[t]
\centering
\includegraphics[width=0.88\linewidth]{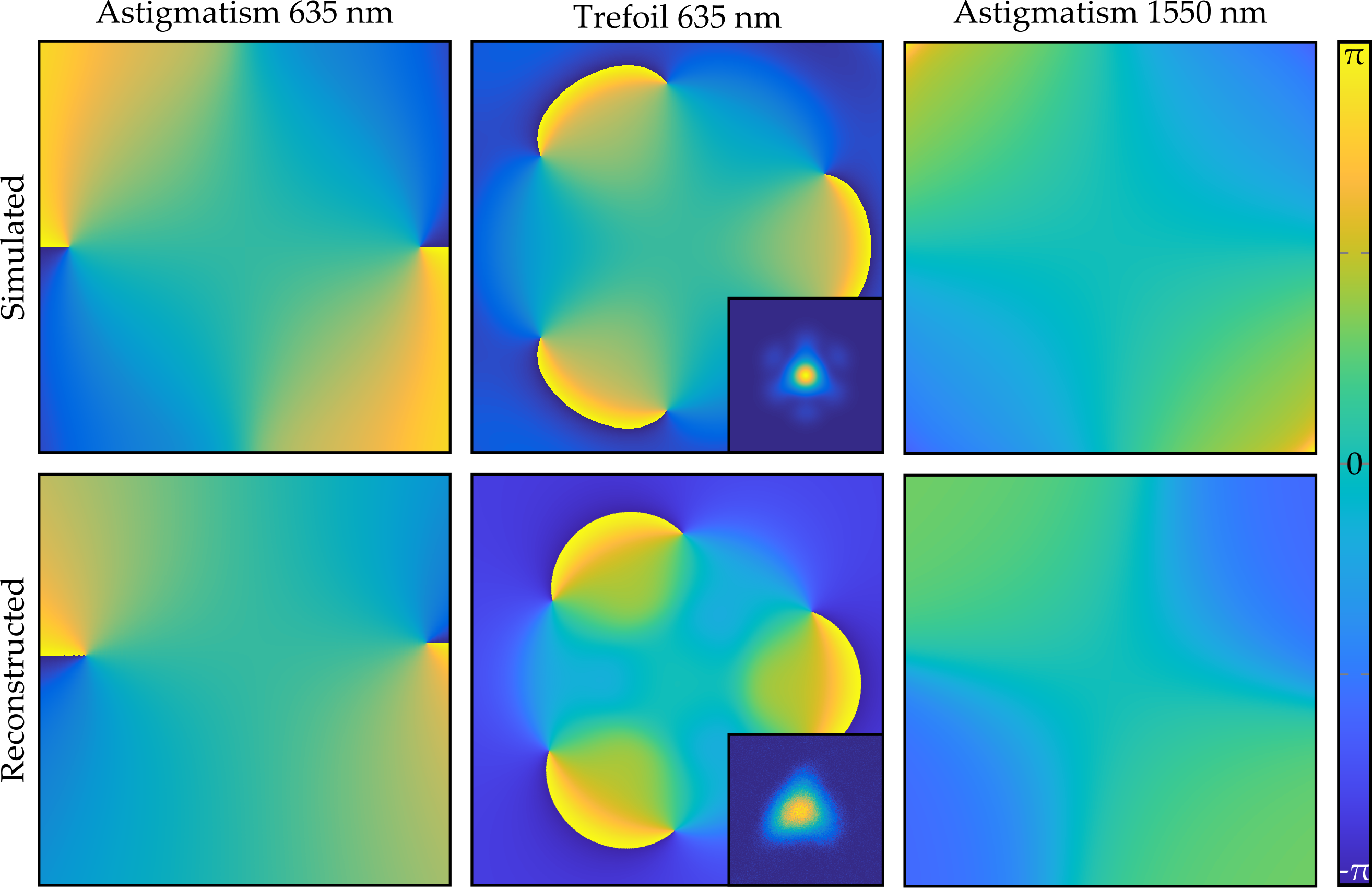}
\caption{Simulated and measured (reconstructed) wavefront measurements for visible and NIR wavelengths of two aberration examples with an inset intensity comparison for the trefoil case. The differences in the intensity of the inset images are due to camera sensitivity.}
\label{fig:visResults}
\end{figure}

Figure~\ref{fig:visResults} shows the reconstruction results for visible wavelengths with astigmatism and trefoil aberrations as examples. For both cases the measured wavefront is remarkably similar to the programmed aberration. A NIR wavefront measurement is shown the the right of Fig.~\ref{fig:visResults}, and is also found to be in excellent agreement with the simulation. The slight difference in ``flatness'' of the measured wavefront with respect to the simulated one was attributed to errors in the inter-modal phase measurements.



\section{Discussion}
For both the visible and NIR tests, the primary cause for inaccuracy is the inter-modal phase measurement. This is consistent with the verification tests in Fig.~\ref{fig:sanityTests}, where the inter-modal phase error was also more prominent than the intensity decomposition error. This is due to noise in the intensity measurements, mainly caused by displacements of the beam during the modal decomposition as a result of air-flow in the laboratory, and to some extent to the compounding of errors in Eq.~(\ref{eq:intermodalPhase}). 

\Note{A simple error analysis reveals that the percentage error in the phase scales as $4 \Delta I/|I_n^{\Psi} - I_n|$, where $I_n^\Psi$ is the signal in the cosine or sin modes, $I_n^{\mathrm{cos}}$ or $I_n^{\mathrm{sin}}$, and $\Delta I$ is the error due to the detector.  Consequently, the phase error will be negligible for modes of reasonable power since $\Delta I$ can be made very small while $I_n$ is high. On the other hand the phase error can be high for modes of low modal power content (small $I_n$).  Fortuitously, our approach by very definition weights the modes according to modal power, so it is the low power modes that are least important in the reconstruction process. The use of a higher resolution ADC will result in more accurate reconstructions since the systematic error component of $\Delta I$ will be reduced. For example, 16 and 24~bit ADCs have dynamic ranges of 96~dB and 145~dB respectively, which corresponds to nano-Watt intensity measurement accuracy for incoming beams in the hundreds of milli-Watt range. Taking this as a typical case we find the percentage error in phase in the order of $\approx 10^{-6}$.  Provided a suitable photodiode is used, these sensitivities are possible for both visible and NIR wavelengths.}

\Note{In addition, the accuracy of the reconstructed wavefront is dependent on the number of modes used for the decomposition and the complexity of the aberration, as described in Sec.~\ref{sec:theory}. A higher-order Zernike aberration requires more modes to reconstruct than a lower-order aberration.  It has been shown that with only a few modes very complex phase structures can be mapped, often requiring fewer than 10 modes \cite{schulze2012wavefront,schulze2013reconstruction,litvin2011poynting,schulze2013measurement}. Further, in many practical applications (such as thermal aberrations of high-power laser beams or optical aberration of delivered beams) only a few lower-order aberrations are required to describe the beam.  This is true even for the case of low to moderate turbulence, where the first few Zernike terms describe most of the observed wavefront-error.  We can understand this by remembering that the rms wavefront error scales with the square of the Zernike coefficients (the sum of the squared coefficients to be precise), so that small coefficients become negligible. However, in very strong scattering media such as tissue or very strong turbulence where scintillation is experienced, we would expect our technique to require many modes for an accurate reconstruction with high error due to low modal powers. Our interest is in real-time analysis for real-time correction, and in such cases correction would be equally problematic.}

\Note{The resolution of the DMD and the size of the incoming beam sets an upper limit to the number of modes that can be tested and for a SLM with $1920\times1080$ resolution, this is on the order of hundreds of modes \cite{Rosales-Guzman2017}. We can expect similar performance from a DMD. The radius of an LG mode is given by $w_0\sqrt{2p + |\ell| + 1}$ and so for instance, with $w_0=0.5$~mm and a DLP3000 DMD which has a minimum dimension of 608 pixels with pitch 7.6~$\mu$m, an LG mode with $\ell=5$ and $p=5$ will fill the DMD.  This is equivalent to more than $60$ modes whereas less than $10$ modes were needed for accurate wavefront reconstruction in this work.}


One of the benefits of our technique is the potential for real-time wavefront reconstruction. A camera was used for the visible measurements and so the decomposition was simply scripted at low speed ($\approx$60~Hz hologram rate) whereas for the NIR tests a photodiode was used which allowed for faster rates. Initial NIR tests were performed in a similar, scripted manner but a test was performed where we loaded the holograms into the DMDs frame buffer and took measurements at the maximum refresh rate of 4~kHz. The results were identical to the ``slow'' scripted version, proving that wavefront measurements can be done quickly using this method.

Given that multiple modal decomposition measurements are required to reconstruct a single wavefront, it is pertinent to elaborate on the achievable wavefront measurement rates of this technique. Different applications require different wavefront measurement rates, for instance, thermal aberrations typically are slowly evolving over time frames of seconds, while moderate atmospheric turbulence changes at rates of 100s of Hz \cite{Greenwood1977}. 

Table~\ref{tab:rates} shows calculated wavefront reconstruction rates (wavefronts per second) for several different mode-sets. The maximum number of measurements required for the approach in this paper is $3N-2$ where $N$ is the total number of modes in the set. We see that even assuming many modes on a low speed device we are able to do wavefront sensing at video frame rates, whereas for realistic mode sets on better devices the rate becomes in the order of 100s to 1000s of Hz, fast enough to be considered real-time in most applications.  A possible future improvement to the measurement algorithm could make use of compressive sensing techniques and a more targeted measurement regime, thus requiring fewer measurements and resulting in even faster wavefront sensing.

\begin{table}[]
\centering
\caption{Resulting wavefront measurement rate (Hz) for different DMD refresh rates and mode-sets. Larger mode sets will result in higher wavefront reconstruction accuracy.}
\label{tab:rates}
\begin{tabular}{r|c|c|c}
 & $\mathrm{LG}_{p\in[0,3]}^{\ell\in[-3,3]}$ & $\mathrm{LG}_{p\in[0,5]}^{\ell\in[-5,5]}$ & $\mathrm{LG}_{p=0}^{\ell\in[-5,5]}$ \\
 \hline
4 kHz & 48 & 20 & 129 \\
9.5 kHz & 115 & 48 & 306 \\
32 kHz & 390 & 163 & 1032
\end{tabular}
\end{table}

\Note{Finally we point out that the advantage of the modal approach to wavefront sensing is that it simultaneously provides all the required information to infer numerous physical properties of the laser beam, including the Poynting vector, orbital angular momentum density, laser beam quality factor ($M^2$), modal structure and so on, making our DMD modal decomposition approach highly versatile.}

\section{Conclusion}
We have demonstrated a fast, broadband and inexpensive wavefront-sensor built around a DMD spatial light modulator. Owning to the employed modal decomposition technique, the resolution of the reconstructed wavefronts is not determined by the resolution of detector, which is a spatially non-resolved photodiode. On the contrary it solely depends on the resolution of the basis functions, which are purely computational. These advantages allow high-resolution wavefront sensing in real-time. We expect that devices based on this novel approach will be invaluable for wavefront sensing of NIR wavelengths where other approaches are either too challenging or too expensive.

\section*{Funding}

EPSRC Centre for Doctoral Training in Intelligent Sensing and Measurement (EP/L016753/1); EPSRC QuantIC (EP/M01326X/1); ERC TWISTS (340507). 



\end{document}